\documentclass[prl,superscriptaddress,onecolumn,preprint]{revtex4}
\usepackage{amsmath,amssymb,amsthm}
\usepackage{xspace}
\usepackage{graphicx} 
\usepackage{latexsym} 
\usepackage[mathscr]{eucal} 
\usepackage{bm}
\usepackage{dcolumn}
\usepackage{color}

\newcommand{\comment}[1]{}

\begin{document}

\title{Experimental undecidability of macroscopic quantumness}
\author{Pawe{\l} \surname{Kurzy\'nski}}
\affiliation{Centre for Quantum Technologies, National University of Singapore, 3 Science Drive 2, 117543 Singapore, Singapore}
\affiliation{Faculty of Physics, Adam Mickiewicz University, Umultowska 85, 61-614 Pozna\'{n}, Poland} 
\author{Akihito \surname{Soeda}}
\affiliation{Centre for Quantum Technologies, National University of Singapore, 3 Science Drive 2, 117543 Singapore, Singapore}
\author{Ravishankar \surname{Ramanathan}}
\affiliation{Centre for Quantum Technologies, National University of Singapore, 3 Science Drive 2, 117543 Singapore, Singapore} 
\author{Andrzej \surname{Grudka}}
\affiliation{Faculty of Physics, Adam Mickiewicz University, Umultowska 85, 61-614 Pozna\'{n}, Poland}
\affiliation{Centre for Quantum Technologies, National University of Singapore, 3 Science Drive 2, 117543 Singapore, Singapore}  
\author{Jayne \surname{Thompson}}
\affiliation{ARC Centre of Excellence for Particle Physics at the Terascale, School of Physics, The University of Melbourne, Victoria 3010, Australia}
\author{Dagomir \surname{Kaszlikowski}}
\email{phykd@nus.edu.sg}
\affiliation{Centre for Quantum Technologies, National University of Singapore, 3 Science Drive 2, 117543 Singapore, Singapore}
\affiliation{Department of Physics, National University of Singapore, 2 Science Drive 3, 117542 Singapore, Singapore}

\maketitle

\textbf{Quantum mechanics marks a radical departure from the classical understanding of Nature, fostering an inherent randomness which forbids a deterministic description; yet the most fundamental departure arises from something different.
As shown by Bell~\cite{Bell} and Kochen-Specker~\cite{KS}, quantum mechanics portrays a picture of the world in which reality loses its objectivity and is in fact created by observation.  Quantum mechanics predicts phenomena which cannot be explained by any theory with objective realism, although our everyday experience supports the hypothesis that macroscopic objects, despite being made of quantum particles, exist independently of the act of observation; in this paper we identify this behavior as classical.  Here we show that this seemingly obvious classical behavior of the macroscopic world cannot be experimentally tested and belongs to the realm of ontology similar to the dispute on the interpretations of quantum mechanics~\cite{Pais,Everett}.  For small systems such as a single photon~\cite{Zeilinger} or a pair~\cite{Aspect}, it has been experimentally proven that a classical description cannot be sustained.  Recently, there have also been experiments that claim to have demonstrated quantum behavior of relatively large objects such as interference of fullerenes~\cite{fullerenes}, the violation of Leggett-Garg inequality in Josephson junction~\cite{LG},  and interference between two condensed clouds of atoms~\cite{BEC}, which suggest that there is no limit to the size of the system on which the quantum-versus-classical question can be tested.  These behaviors, however, are not sufficient to refute classical description in the sense of objective reality.  Our findings show that once we reach the regime where an Avogadro number of particles is present, the quantum-versus-classical question cannot be answered experimentally.}

\begin{figure}[t]
\begin{center}
\includegraphics[scale=0.7]{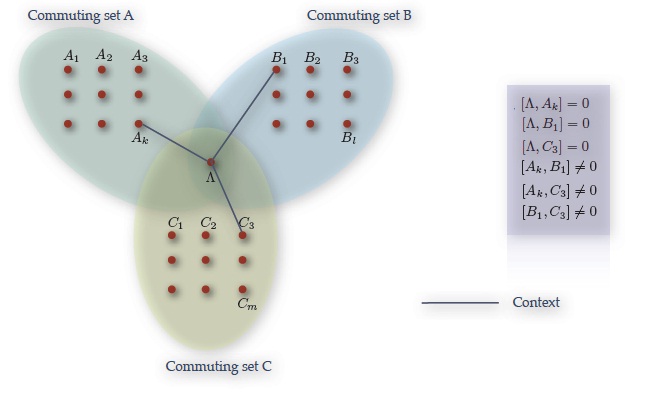}
\end{center}
\caption{Each ellipsoid denotes a set of mutually commuting observables. The observable $\Lambda$ belongs to all three sets. Observables in different sets do not commute. The observable $\Lambda$ can be measured in three different contexts, for instance it can be measured either with $A_k$ or with $B_1$ or with $C_3$. We say that the three observables $A_k, B_1,C_3$ provide contexts for the measurement of $\Lambda$. Note that the proof of KS requires both commutativity that provides contexts and non-commutativity that makes it impossible to perform simultaneous measurements in different contexts.}
 \label{figcontext}
\end{figure}

There is only one known experimental test capable of distinguishing between classical (objective realistic) theories and non-classical ones.  This test is based on the celebrated paper by Kochen and Specker~\cite{KS} and exploits a simple observation that is at the root of the discrepancy between classical and non-classical behavior.  Imagine a physical system on which a set of $N$ measurements can be performed that are represented by some physical observables $A_1,A_2,\dots, A_N$.  Each observable yields outcomes of measurements $a^{(i)}_j$ with probability distribution $p(a^{(i)}_j)$.  The physical system is classical if and only if there exists a joint probability distribution of the outcomes of measurements for all involved observables $p(a^{(1)}_j,a^{(2)}_{j'},\dots,a^{(N)}_{j''})$~\cite{Fine}.  Indeed, known classical theories such as Maxwell electrodynamics and general relativity satisfy this property.  One can immediately have doubts whether this is true in quantum theory because some observables in this theory are incompatible.  For instance, observables position $X$ and momentum $P$ cannot be measured simultaneously as is neatly expressed by the Heisenberg uncertainty principle.  However, this does not prevent one from constructing a joint probability distribution for $X$ and $P$ as $Prob(X=x,P=p)=Prob(X=x)Prob(P=p)$.  This probability distribution directly reproduces observed marginal probabilities for both position and momentum measurements $Prob(X=x)=\int dp Prob (X=x,P=p), Prob(P=p)=\int dx Prob(X=x,P=p)$.   This simple example teaches us that we need more than incompatibility of measurements to demonstrate the non-existence of a joint probability distribution.  We require measurements that can be performed in different \textit{contexts}, the simplest example being an observable $A$ that can be measured together with an observable $B$ or with $C$ such that $B$ and $C$ cannot be measured simultaneously.  In quantum mechanics, this is written as $[A,B]=[A,C]=0$ and $[B,C] \neq 0$, where $B$ and $C$ are understood to provide two different contexts for the measurement of $A$ (Fig.~\ref{figcontext}).  In any non-contextual theory, the outcome of the measurement of $A$ does not depend on whether it was co-measured with $B$ or with $C$.   To show that quantum theory is a contextual theory, i.e., it does not allow for a joint probability distribution, one needs more than two contexts as shown in \cite{KS}.  For a spin-1 system, the lack of the joint probability distribution has been experimentally confirmed in \cite{Zeilinger, Kl}.  If one wants to perform a similar experiment on a macroscopic system it is necessary to find a set of observables such that some of them commute and some do not.  

However in the macroscopic regime, the class of measurements that can be performed is limited to a small set of feasible measurements.  For simplicity, let us consider a magnetic system consisting of $N$ spin-$s$ particles.  Despite the generalized spin measurement proposal in \cite{GenSG}, when the number of spins becomes too large ($N \approx 10^{23}$), feasible measurements will be limited to measurements of magnetization in some direction $\vec{n}$.  These are given by a set of projectors of the form
\begin{eqnarray}
&&P_m(\vec{n})=\sum_{k_1+\dots + k_N=m}P_{k_1}(\vec{n})\otimes\dots\otimes P_{k_N}(\vec{n}), \label{e1}
\end{eqnarray}
where each $k_i=-s,-s+1,\dots s$ and $m$ corresponds to different degrees of magnetization.  This limitation on the set of measurements that can be performed is imposed by the following two factors. 

The first restriction is imposed by the fact that a physical description for a system consisting of a large number of particles $N$ can only be done by statistical theories.  The basic assumption in these theories is that one cannot know the exact micro-state of the system and is hence forced to assume \textit{a priori} that all micro-states leading to the same macro-state are equally probable.  One of the consequences of this assumption is that any observable that can be measured must not distinguish between micro-states having the same macroscopic property.  Indeed, all states corresponding to a given property belong to a permutationally invariant subspace; this in turn implies effective indistinguishability of the particles and yields the form of the projectors in equation (\ref{e1}). This permutational invariance of the projectors is therefore a key feature of feasible measurements on any macroscopic system. 

The second restriction is due to the fact that, for spin systems, magnetization dominates more exotic multipole moments.  In principle it is possible to measure the $k$-pole moment of magnetization for a spin-$s$ particle, where $k\leq 2s+1$~\cite{GenSG}.  Therefore, there is no theoretical limitation to measuring a macroscopic $k$-pole observable that, analogous to magnetization, is a sum of $k$-pole moments of all spins in the system.  However, the electromagnetic fields required for measuring the $k$-pole moments are infeasible to implement in practice.  Moreover, higher moments such as susceptibility are not suitable observables to study contextuality as they deal with states that are varying in time.  Hence, the feasible measurements for macroscopic spin systems are restricted to magnetization.

The main result of this work is that for macroscopic systems the question of quantum-versus-classical is undecidable, i.e., is not experimentally testable.  This is because for macroscopic measurements there is no context.  We show that $[P_m(\vec{n}),P_{m'}(\vec{n}')]=0$ if and only if $\vec{n}=\pm \vec{n}'$ (Fig.~\ref{figvect}).  Details of the proof can be found in the supplementary material.   The proof holds mainly due to the fact that the restrictions on the feasible measurements force the measurable obeservables to be related by three-dimensional spatial rotations instead of much more complex rotations in the Hilbert space.   This undecidability sheds new light on the quantum-versus-classical question suggesting that it may be forever unanswerable.

\begin{figure}[t]
\begin{center}
\includegraphics[scale=0.7]{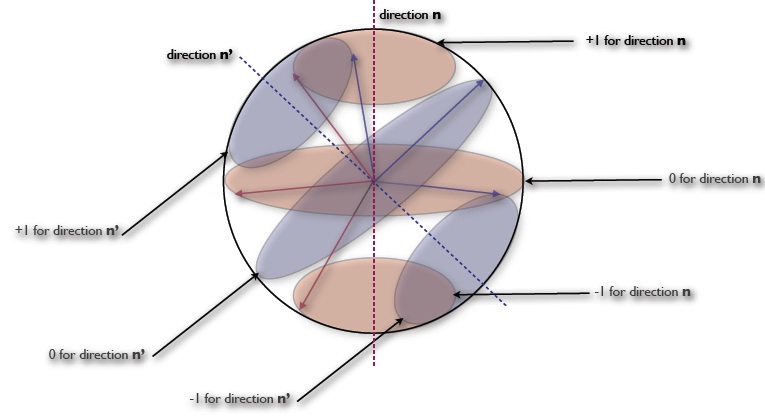}
\end{center}
\caption{Graphical representation of spin-1 (the smallest contextual system) states along two different directions $\vec{n},\vec{n}'$.  Spin-1 can be represented as a vector of length $\sqrt{2}$ in three dimensional space and has three values along any measurement direction, $+1,0,-1$, therefore we have three arrows per direction. Due to the Heisenberg uncertainty only one coordinate of the spin vector can be determined, namely the one along the measurement direction. The other two coordinates remain undetermined hence for each measurement outcome along direction $\vec{n}$ (or $\vec{n}'$) the tip of the spin vector is spread over the rim of the corresponding circle. Red circles depict the projectors $P_{-1}(\vec{n}),P_{0}(\vec{n}),P_{+1}(\vec{n})$ and blue ones depict $P_{-1}(\vec{n}'),P_{0}(\vec{n}'),P_{+1}(\vec{n}')$.  $[P_m(\vec{n}),P_{m'}(\vec{n}')]=0$ if and only if the two vectors are parallel.  Similar results hold for systems composed of a large number of spins of any dimension.}
 \label{figvect}
\end{figure}

Interestingly, it was shown in~\cite{macro-bell} that the quantumness of the correlations between composite macroscopic objects is experimentally testable and conforms with (local) realism.  This situation resembles the case of a single qubit versus a pair of qubits.  The quantumness of a single qubit is fundamentally undecidable~\cite{Gleason,Mermin} because no contexts can be found for any measurements, whereas the correlations between two qubits are decidable and for entangled states can be shown to be quantum~\cite{Aspect}.

It is important to note that the various notions of quantumness, such as interference and the violation of Leggett-Garg inequality~\cite{fullerenes,LG,BEC}, have been introduced in the literature.  The experiments mentioned in the introductory paragraph may confirm the quantumness of physical systems according to one of these notions.  They do not, however, involve measurements of different contexts and hence do not refute the existence of objective reality.

To conclude, we showed that the restricted set of measurements in macroscopic systems leads to a simplification of quantum theory that prevents the testability of certain non-intuitive features such as contextuality while still preserving other features such as entanglement \cite{macro-bell, polzik}.  Similar behavior was observed in a toy model of quantum theory proposed in \cite{spekkens}. An interesting open question is to investigate the complexity of measurements needed for contextuality to emerge.  A further question could be to identify the exact trade-off between the size of the system and the complexity of the measurements that can be performed on it. 

\textit{Acknowledgments} This work is supported by the National Research Foundation and Ministry of Education in Singapore. A.G. acknowledges the support of the Polish Ministry of Science and Higher Education grant nr IdP2011 000361.

\section{Appendix}

Here we prove that for measurements of macroscopic magnetizations along two nonparallel directions all projectors corresponding to one magnetization measurement do not commute with all projectors corresponding to the other magnetization measurement. Let us denote projectors of the first measurement, that defines the Z axis, as $\mathbb{P}_m$, where $m$ labels eigenvalues of magnetization operator along Z. The projectors of the second measurement (along a direction that defines the XZ plane) are denoted by $\tilde{\mathbb{P}}_{n}$ and are related to $\mathbb{P}_n$ via rotation operator $\mathcal{D}(\beta)$, i.e., $\tilde{\mathbb{P}}_{n}=\mathcal{D}(\beta)\mathbb{P}_{n} \mathcal{D}^{\dagger}(\beta)$. Since we are dealing with $N\gg 1$ particles, there are two different representations of $\mathbb{P}_m$. The first one is given via the tensor product of spin states corresponding to different particles
\begin{equation}
\mathbb{P}_m=\sum_{k_1+\dots+k_N=m}\left(|k_1\rangle\otimes\dots\otimes|k_N\rangle\right)\left( \langle k_1|\otimes\dots\otimes \langle k_N|\right),
\end{equation} 
whereas the second is given via a direct sum of spin states corresponding to different values of total angular momentum
\begin{equation}\label{e2}
\mathbb{P}_m=\sum_{j,\lambda_j}|j,\lambda_j,m\rangle\langle j,\lambda_j,m|,
\end{equation}
where $j$ runs over all values allowed by standard angular momenta addition rules and $\lambda_j$ denotes the degeneracy of $j$, i.e., the number of different realizations of spin $j$ with $N$ particles. Note, that $N \gg 1$ leads to a high degeneracy of $\mathbb{P}_m$ and as a result guarantees that the sum in (\ref{e2}) runs over many different values of $j$. In this proof we exploit the properties of the second representation. 

The general form of the rotation matrix (often referred to as the Wigner d-matrix) is the following
\begin{equation}
\mathcal{D}(\beta)=\sum_{j,\lambda_{j},m,\tilde{m}}d^{j,\lambda_j}_{\tilde{m},m}(\beta)|j,\lambda_j,\tilde{m}\rangle\langle j,\lambda_j,m|.
\end{equation}
It follows that
\begin{equation}
\tilde{\mathbb{P}}_{n}=\sum_{j,\lambda_{j},\tilde{m},\tilde{m'}}d^{j,\lambda_j}_{\tilde{m},n}(\beta)d^{j,\lambda_j}_{\tilde{m'},n}(\beta)|j,\lambda_j,\tilde{m}\rangle\langle j,\lambda_j,\tilde{m'}|,
\end{equation}
and
\begin{equation}
[\mathbb{P}_m,\tilde{\mathbb{P}}_{n}]=\sum_{j,\lambda_{j},\tilde{m}}d^{j,\lambda_j}_{m,n}(\beta)d^{j,\lambda_j}_{\tilde{m},n}(\beta)\left(|j,\lambda_j,m\rangle\langle j,\lambda_j,\tilde{m}|-|j,\lambda_j,\tilde{m}\rangle\langle j,\lambda_j,m|\right).
\end{equation}
The above commutator equals zero if all terms of the corresponding matrix vanish. Let us define
\begin{equation}
\Gamma_{k,k'}^{j,\lambda_{j}}=\langle j,\lambda_j,k|[\mathbb{P}_m,\tilde{\mathbb{P}}_{n}]|j,\lambda_j,k'\rangle.
\end{equation}
In the above we do not consider off-diagonal terms corresponding to different $j$'s and $\lambda$'s, since they are trivially equal to zero. One easily finds
\begin{equation}
\Gamma_{k,k'}^{j,\lambda_{j}}=d^{j,\lambda_j}_{m,n}(\beta)\left(\delta_{k,m}d^{j,\lambda_j}_{k',n}(\beta) - \delta_{k',m}d^{j,\lambda_j}_{k,n}(\beta)\right).
\end{equation}

The commutator vanishes if $\Gamma_{k,k'}^{j,\lambda_{j}}=0$ for all allowed $j,\lambda_j,k,k'$. Since there is no dependency on $\lambda_j$, from now on we skip this superscript. Below we show that there always exist a set of $j,k,k'$ for which $\Gamma_{k,k'}^{j}\neq 0$. First, let us note that $\Gamma_{k,k'}^{j}$ can be nonzero only for $k=m$ and $k'\neq m$, or for $k\neq m$ and $k'=m$. Effectively, without loosing generality, it is enough to show that there exist $j$ and $k \neq m$ for which 
\begin{equation}
d^{j}_{m,n}(\beta)d^{j}_{k,n}(\beta)\neq 0. 
\end{equation}
At this stage let us write explicitly
\begin{eqnarray}
d^{j}_{m',m}(\beta)&=&\sqrt{\frac{(j+m')!(j-m')!}{(j+m)!(j-m)!}}\left(\cos\frac{\beta}{2}\right)^{-m-m'}\left(\sin\frac{\beta}{2}\right)^{2j+m+m'}(-1)^{j+m'} \times \nonumber \\ &\times& \sum_{\nu}\left((-1)^{\nu} \begin{pmatrix} j+m \\ \nu \end{pmatrix} \begin{pmatrix} j-m \\ j+m' -\nu \end{pmatrix} \left(\cos\frac{\beta}{2}\right)^{2\nu} \left(\sin\frac{\beta}{2}\right)^{-2\nu} \right), \label{d}
\end{eqnarray}
where $\nu$ goes over all integer values $\nu \geq 0$ for which the binomial coefficients do not vanish. 

We are going to find conditions under which $d^{j}_{m',m}(\beta)$ is nonzero. Let us note that in case of two nonparallel magnetization directions ($0<\beta <\pi$) all trigonometric functions in the above formula are nonzero. Let us consider two cases. First, let us assume that both $m'=m=0$. In such a case
\begin{equation}
d^{j}_{0,0}(\beta)=P_j(\cos\beta),
\end{equation}
where $P_j(x)$ is the Legendre polynomial that obeys
\begin{equation}\label{rel}
\sum_{j=0}^{\infty}P_j(x)t^j=\frac{1}{\sqrt{1-2xt+t^2}},
\end{equation}
and the recursion relation
\begin{equation}
(j+1)P_{j+1}(x)=(2j+1)x P_j(x)- j P_{j-1}(x).
\end{equation}
We claim that in the allowed range of $j$ for any $-1<x<1$ there exists $j$ for which $P_j(x) \neq 0$. Otherwise, recursion relation would imply that $P_j(x) \neq 0$ for at most a finite set of $j$, but this is inconsistent with (\ref{rel}), hence it is not possible. Next, consider a case when at least one of labels ($m$ or $m'$) differs from zero. In this case we set $j=\max\{|m|,|m'|\}$ and it is easy to see that the sum in (\ref{d}) has only one nonzero term.

Finally, let us show that the product of two coefficients $d^{j}_{m,n}(\beta)d^{j}_{k,n}(\beta)$ differs from zero. Since in this case $j$ is the same for both coefficients and $m$ and $n$ are fixed, let us again consider two cases. Following the above discussion, if both $m=n=0$ we set $j$ such that the first coefficient is nonzero and then we set $k=j$. On the other hand, if $m$ or $n$ are nonzero we can guarantee that the first coefficient does not vanish by fixing $j=\max\{|m|,|n|\}$. For the second coefficient note that $k\neq m$. In case $|m|\leq|n|$ it is possible to set $k$ such that $|k|\leq |n|$
and $k\neq m$. Since  $j=\max\{|m|,|n|\}=|n|$ both coefficients are nonzero. In case $|m|>|n|$ we set $k=-m$ and $j=\max\{|m|,|n|\}=|m|$ prevents both coefficients from being zero. This ends the proof.


\begin{thebibliography}{99}

\bibitem{Bell}
J. S. Bell, 
Physics {\bf 1}, 195 (1964).

\bibitem{KS}
S. Kochen and E.P. Specker,
J. Math. Mech. {\bf 17}, 59 (1967).

\bibitem{Pais}
 A. Pais,
Rev. Mod. Phys. {\bf 51}, 863 (1979).

\bibitem{Everett}
H. Everett,
Rev. Mod. Phys. {\bf 29}, 454 (1957).


\bibitem{Zeilinger}
R. Lapkiewicz {\it et al.}, 
Nature (London) {\bf 474}, 490 (2011).

\bibitem{Aspect}
A. Aspect, J. Dalibard, and G. Roger, 
Phys. Rev. Lett. {\bf 49}, 1804 (1982).

\bibitem{fullerenes}
M. Arndt {\it et al.},
Nature (London) {\bf 401}, 680 (1999).

\bibitem{LG}
A. Palacios-Laloy {\it et al.},
Nat. Phys. {\bf 6}, 442 (2010).

\bibitem{BEC}
M. R. Andrews {\it et al.}, 
Science {\bf 275}, 637 (1997).

\bibitem{Fine}
A. Fine,
Phys. Rev. Lett. {\bf 48}, 291 (1982).

\bibitem{Kl}
A. A. Klyachko {\it et al.}, 
Phys. Rev. Lett. {\bf 101}, 020403 (2008).

\bibitem{GenSG}
A. R. Swift and R. Wright,
J. Math. Phys. {\bf 21}, 77 (1980).

\bibitem{macro-bell}
R. Ramanathan {\it et al.},
Phys. Rev. Lett. {\bf 107}, 060405 (2011).

\bibitem{Gleason}
A. M. Gleason, 
J. Math. Mech. {\bf 6}, 885 (1957).

\bibitem{Mermin}
N. D. Mermin, 
Rev. Mod. Phys. {\bf 65}, 803 (1993).


\bibitem{polzik}
B. Julsgaard {\it et al.}, 
Nature (London) {\bf 413}, 400 (2001).

\bibitem{spekkens}
R. W. Spekkens,
Phys. Rev. A {\bf 75}, 032110 (2007) 



\end{thebibliography}
\end{document}